\documentclass[aps,prc,twocolumn,superscriptaddress]{revtex4-1}

\usepackage{lineno}
\usepackage{graphicx}
\usepackage{amssymb}
\usepackage{amsmath}
\usepackage{float}

\begin{document}



\title{Single-reference high-precision mass measurement with\\a multi-reflection time-of-flight mass spectrograph}


\author{Y. Ito}
\affiliation{SLOWRI Team, Nishina Accelerator-based Research Center, RIKEN, 2-1 Hirosawa, Wako, Saitama 351-0198, Japan}
\affiliation{University of Tsukuba, 1-1-1 Tennodai, Tsukuba, Ibaraki 305-8577, Japan}
\author{P. Schury}
\affiliation{SLOWRI Team, Nishina Accelerator-based Research Center, RIKEN, 2-1 Hirosawa, Wako, Saitama 351-0198, Japan}
\affiliation{University of Tsukuba, 1-1-1 Tennodai, Tsukuba, Ibaraki 305-8577, Japan}
\affiliation{New Mexico State University, Department of Chemistry and Biochemistry, Las Cruces, New Mexico 88003,USA}
\author{M. Wada}
\affiliation{SLOWRI Team, Nishina Accelerator-based Research Center, RIKEN, 2-1 Hirosawa, Wako, Saitama 351-0198, Japan}
\author{S. Naimi}
\affiliation{SLOWRI Team, Nishina Accelerator-based Research Center, RIKEN, 2-1 Hirosawa, Wako, Saitama 351-0198, Japan}
\author{T. Sonoda}
\affiliation{SLOWRI Team, Nishina Accelerator-based Research Center, RIKEN, 2-1 Hirosawa, Wako, Saitama 351-0198, Japan}
\author{H. Mita}
\affiliation{SLOWRI Team, Nishina Accelerator-based Research Center, RIKEN, 2-1 Hirosawa, Wako, Saitama 351-0198, Japan}
\affiliation{University of Tsukuba, 1-1-1 Tennodai, Tsukuba, Ibaraki 305-8577, Japan}
\author{F. Arai}
\affiliation{SLOWRI Team, Nishina Accelerator-based Research Center, RIKEN, 2-1 Hirosawa, Wako, Saitama 351-0198, Japan}
\affiliation{University of Tsukuba, 1-1-1 Tennodai, Tsukuba, Ibaraki 305-8577, Japan}
\author{A. Takamine}
\affiliation{SLOWRI Team, Nishina Accelerator-based Research Center, RIKEN, 2-1 Hirosawa, Wako, Saitama 351-0198, Japan}
\affiliation{College of Science and Engineering, Aoyama-Gakuin University, 5-10-1 Fuchinobe, Sagamihara, Kanagawa 252-5258, Japan}
\author{K. Okada}
\affiliation{SLOWRI Team, Nishina Accelerator-based Research Center, RIKEN, 2-1 Hirosawa, Wako, Saitama 351-0198, Japan}
\affiliation{Sophia University, 7-1 Kioi-cho, Chiyoda-ku, Tokyo 102-8554, Japan}
\author{A. Ozawa}
\affiliation{University of Tsukuba, 1-1-1 Tennodai, Tsukuba, Ibaraki 305-8577, Japan}
\author{H. Wollnik}
\affiliation{SLOWRI Team, Nishina Accelerator-based Research Center, RIKEN, 2-1 Hirosawa, Wako, Saitama 351-0198, Japan}
\affiliation{New Mexico State University, Department of Chemistry and Biochemistry, Las Cruces, New Mexico 88003,USA}


\date{\today}

\begin{abstract}
\par A multi-reflection time-of-flight mass spectrograph, competitive with Penning trap mass spectrometers, has been built at RIKEN.
We have performed a first online mass measurement, using $^{8}$Li$^{+}$ ($T_{1/2} = 838$ ms).
A new analysis method has been realized, with which, using only $^{12}$C$^+$ references, the mass excess of $^{8}$Li was accurately determined to be 20\,947.6(15)(34) keV ($\delta m/m = 6.6 \times 10^{-7}$).
The speed, precision and accuracy of this first online measurement exemplifies the potential for using this new type of mass spectrograph for precision measurements of short-lived nuclei.
\end{abstract}


\maketitle

\par Mass measurements of unstable nuclei, providing direct measure of the nuclear binding energy, are invaluable for nuclear structure and nuclear astrophysics.
Mass measurements of highly neutron-rich nuclei from Co to Xe, of importance for understanding both the astrophysical {\it r} process and evolution of shell structure, require fast measurement time and high efficiency, due to their typically short lifetimes ($T_{1/2}<100\ \rm{ms}$) and low production yields.
\par The most precise atomic mass measurements are obtained from Penning trap mass spectrometry (PTMS) of stable nuclei.
The observation time required for PTMS to achieve a given resolving power, $R_{m}$ is
\begin{equation}
 t_{\rm obs} = m R_{m}/qB,
 \label{eq:PTMS_t_obs}
\end{equation}
where $m/q$ is the mass-to-charge ratio, $B$ is the Penning trap magnetic field strength and $t_{\rm obs}$ is the observation time in the trap  \cite{Blaum2006}.
While PTMS can achieve resolving powers of several million, doing so requires $t_{\rm obs}\gtrsim 100$~ms.
The linear scaling of $t_{\rm obs}$ with the mass-to-charge ratio limits the maximum resolving power that can be achieved for short-lived, heavy nuclei.
A few methods to mitigate this limitation have been considered -- using higher magnetic fields  \cite{Ringle2005}, charge breeding  \cite{Ettenauer2011} and higher-order multipole excitation  \cite{Ringle2007,Eliseev2011}.
We think that a separate path may prove more fruitful. 
\par By using a pair of electrostatic mirrors \cite{Wollnik1990}, the flight path for a pulse of ions, e.g., from an ion trap, could be extended indefinitely.
The time, $t$, required for ions to travel to a detector on the far side of the mirrors can be written as 
\begin{equation}
 t \approx \sqrt{m} \int^L_0\left(dx/\sqrt{K(x)}\right),
 \label{eq:MRTOF_t}
\end{equation}
where $K$ is the ion's kinetic energy and $L$ is the total flight length.
The mass resolving power of such a measurement is simply
\begin{equation}
 R_{m} = \tfrac{1}{2}R_{t} = \tfrac{1}{2}t/\Delta t,
 \label{eq:mtof_resolution_001}
\end{equation}
where $R_{t}$ is the time resolving power and $\Delta t$ is the detected pulse width.
Using the electrostatic mirrors to achieve an energy isochronous time-focus at the detector, it is possible to achieve conditions wherein $\Delta t \propto\sqrt{m}$ is completely determined by conditions (e.g., ion temperature) in the ion trap.
By making $\Delta t$ sufficiently small and $t$ sufficiently long, it is possible to achieve reasonably large resolving powers faster than could be achieved by PTMS, i.e., achieve higher resolving powers for sufficiently short-lived nuclei than could be achieved by PTMS.
Comparing the mass dependencies of Eqs. \ref{eq:PTMS_t_obs} and \ref{eq:MRTOF_t}, one can immediately see that this effect becomes ever more pronounced as the mass-to-charge ratio increases.
Thus, there is great potential for using this new type of mass spectrograph for precision measurements of heavy, short-lived nuclei.
\par At RIKEN we have developed such a multi-reflection time-of-flight mass spectrograph (MRTOF) \cite{Ishida2004, Ishida2005, Naimi2013} as part of the SLOWRI facility for low-energy nuclear physics at RIKEN \cite{Wada2003}.
Similar developments at various other facilities have been made with a purpose of isobar purification \cite{Piechaczek20084510,Plass20084560,Wolf201282}.
We will use it to measure the masses of r-process nuclei created by in-flight fission of uranium at the BigRIPS facility \cite{Kubo2003, Ohnishi2010} and trans-uranium elements created by fusion reactions at the GARIS facility \cite{Morita2004, Morita2012}.
In both cases, it is desirable to achieve relative mass precision of $\delta m/m < 10^{-6}$ for heavy (80 $< A/q <$ 280) nuclei with short lifetimes (5 ms $< T_{1/2} <$ 100 ms).
\par As the radioactive nuclei are produced at high energies, they must be thermalized in a helium-filled gas cell to convert them to a low-energy ion beam amenable to such mass measurements.
Prior to constructing a gas cell suited for such heavy nuclei, an already existent gas cell -- designed for use in ion trap laser spectroscopy of Be isotopes \cite{Takamine2005, Okada2008} -- was used to thermalize $^8$Li ions as an initial online test of the MRTOF.
\par
\begin{figure}[t]
 \centering
  \includegraphics[bb=14 5 808 419, width=86mm]{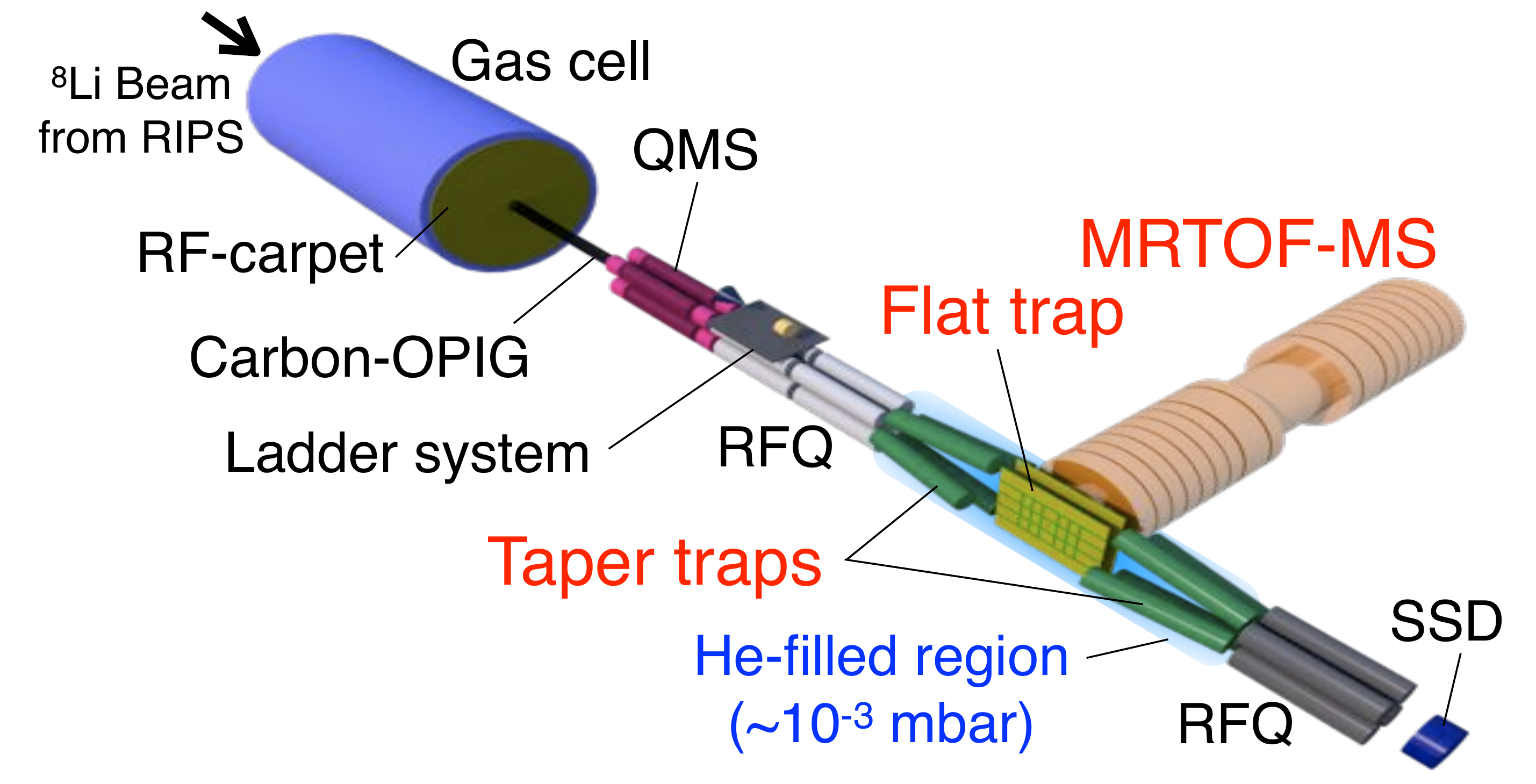}
  \caption{(Color online).
  Sketch of the experimental setup (not to scale).
  A section of RFQ between the QMS and the taper trap was mounted on a ladder system along with a channeltron electron multiplier, which can be used as a beam intensity monitor, and an alkali ion source for offline test.
  An SSD detector was used as the $\alpha$ decay detector of $^{8}$Li for beam line tuning.
  }
  \label{fig:gc_mtof_002}
\end{figure}

 The online experimental setup consisted of a gas cell filled with 20 mbar He gas, rf-multipole transport system, buffer gas-filled ion trap, MRTOF, offline ion source and detector suite, as shown in Fig.~\ref{fig:gc_mtof_002}.
$^{8}$Li ions were produced by projectile fragmentation of a 100$A$ MeV primary beam of $^{13}$C on a 1.86 g/cm$^{2}$ Be production target, selected by the RIPS projectile fragment separator  \cite{Kubo1992} and transported to the prototype SLOWRI branch, where the high-energy $^{8}$Li ions were first decelerated by a wedge-shaped energy degrader before being stopped in the gas cell.
The $^{8}$Li ions were extracted as $^{8}$Li$^+$ by an rf-carpet system \cite{Wada2003} and transported to high vacuum by an octupole ion guide (OPIG) made of resistive carbon fiber reinforced plastic to allow simple production of an axial drag force \cite{Takamine2007}.
Stable ions produced in the gas cell could be largely eliminated by a quadrupole mass separator (QMS).
The $A/q=8$ beam comprised $^{8}$Li$^{+}$ and a small amount of $^{2}$He$_{2}^{+}$ produced in the gas cell.
\par The efficiency from a 1 GeV $^{8}$Li beam to a continuous 5~eV beam by a 2-m-long gas cell with an rf-carpet ion guide was $\approx$ 5\% in Ref.~\cite{Takamine2005} and the MRTOF efficiency from the 5 eV continuous beam was 2.7\%.
Thus, the total efficiency was $\approx$ 0.14\%.
In general, higher $Z$ ions yield higher stopping efficiency, gas-cell extraction efficiency and trapping efficiency.
In an offline test with $^{23}$Na ions, the MRTOF efficiency was found to be 13\%.
\par Prior to being analyzed by the MRTOF, ions are prepared in a sequential pair of buffer gas-filled rf ion traps \cite{Ito2013}.
Ions are initially stored and precooled in a linear Paul trap built with tilted rods (taper trap) before being transferred to a novel ``flat trap".
The flat trap quickly cools ions to a very small cloud and then ejects them toward the MRTOF by means of an electric dipole field \cite{Schury2009}.
The small ion cloud and nearly pure dipole extraction field provide ideal conditions for analysis with the MRTOF. 
\par The taper trap consists of four rod electrodes arranged with small angles to the centerline.
The vertical inter-rod gap at the exit side (i.e., flat trap side) is slightly larger than that at the entrance side, while the horizontal inter-rod gap is slightly smaller.
The asymmetric configuration generates a potential gradient to push the ions toward the flat trap when dc biases are superimposed on the rf signals \cite{Mansoori1998}.
\par The flat trap is built from a pair of flat printed circuit boards mounted in an aluminum frame, separated from each other by 4 mm.
Each circuit board consists of three strips divided into seven segments.
The central electrode of each board has a plated hole with a diameter of 0.8~mm at its center to allow orthogonal extraction of the ion bunch.
While a traditional Paul trap creates a well-approximated quadrupole field using four rod electrodes, the flat trap design approximates a quadrupole field using six strip electrodes.
Complementary pulses of $\pm$60 V applied to the central electrodes create a dipole electric field for orthogonally extracting ions from the trap.
\par
\begin{figure}[t]
 \centering
  \includegraphics[bb=10 41 952 807, width=86mm]{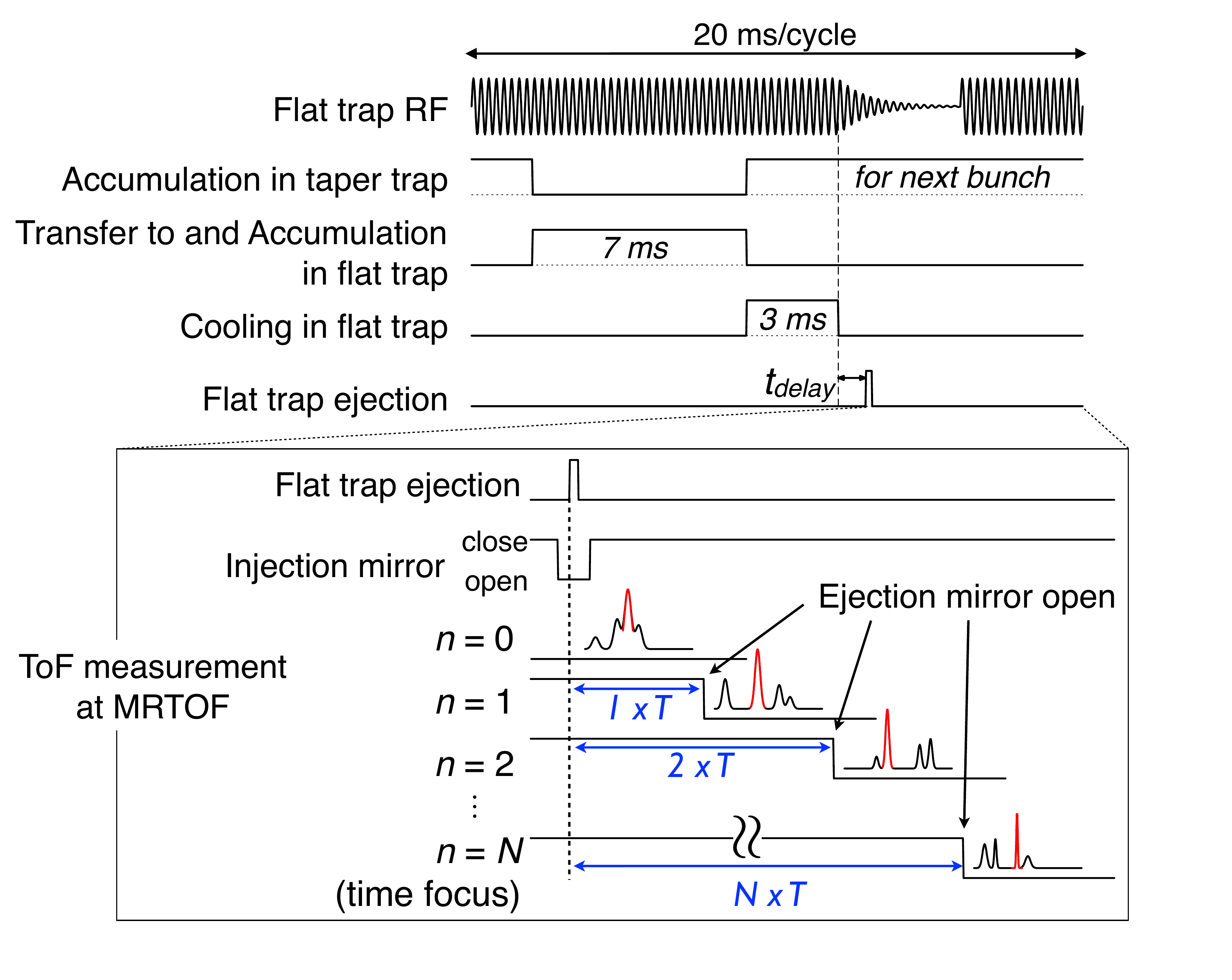}
  \caption{(Color online).
  Measurement time sequence for the trap and MRTOF system (not to scale).
  After the species of interest, indicated by the red peak, makes $N$ laps, the potential of the MRTOF ejection mirror is reduced to allow ions passage and the ions are detected with a microchannel plate.
  Only ions with $A/q$ sufficiently similar to the species of interest make $N$ laps, while heavier (lighter) ions make fewer (more) laps.
  }
  \label{fig:time_sequence_001}
\end{figure}

 During the measurement of $^8$Li$^+$, the taper trap was utilized as an auxiliary trap to accumulate and precool ions while an ion bunch was cooling in the flat trap.
As shown in Fig.~\ref{fig:time_sequence_001}, ions were accumulated and precooled in the taper trap for 10 ms before being transferred to the flat trap.
To maximize the accumulation efficiency, 7 ms was allotted for the transfer to and accumulation in the flat trap.
After accumulating the ions in the flat trap, the trap axial potential well was deepened for 3 ms to maximally cool the ion cloud prior to ejection to the MRTOF.
While one ion bunch is cooling in the flat trap, the next bunch is already accumulating in the taper trap, thereby, allowing an operational duty cycle of $\approx$ 100\%.
\par After cooling, the flat trap rf signal was briefly turned off to minimize the effect of the rf field on ions leaving the trap.
Due to the high {\it Q}-value of the rf resonant circuit, however, the rf amplitude decayed exponentially with a decay constant of $\tau \approx$ 250 $\mu$s.
Since waiting for the amplitude to fully decay would allow the ion cloud to expand, the ejection was instead phase locked to a point in the amplitude decay found to yield a maximum resolving power.
The phase locked ejection signal served as the TDC start signal for the ion time-of-flight (ToF).
\par Prior to ejection of ions from the flat trap, the potential of the first MRTOF injection mirror electrode was reduced by 1 kV to allow the ions entry.
The potential was then returned to its nominal value at the time when the ions were in the MRTOF ejection mirror electrodes, to minimize any effects from the changing potential.
The ions were then allowed to reflect between the two mirrors for a time sufficient to allow the ions to make 880 laps.
After 880 laps, while the ions are in the injection mirror, the final ejection mirror electrode potential was reduced by 1 kV to allow ions to exit and travel to a multi-channel plate detector, providing stop signals for the TDC.
The measurement time sequence is shown in Fig.~\ref{fig:time_sequence_001}.
\par As in any other mass spectroscopic technique, reference measurements are required to determine the mass from the time of flight.
Ideally, isobaric references would be used.
In the case of $^8$Li$^+$, however, the only isobaric reference available was $^4$He$_2^+$, the rate of which was almost an order of magnitude less than that of $^8$Li$^+$.
As such, $^{12}$C$^+$ from the gas cell were used as a reference.
\par
\begin{figure}[t]
 \centering
  \includegraphics[bb=20 71 721 783, width=76mm]{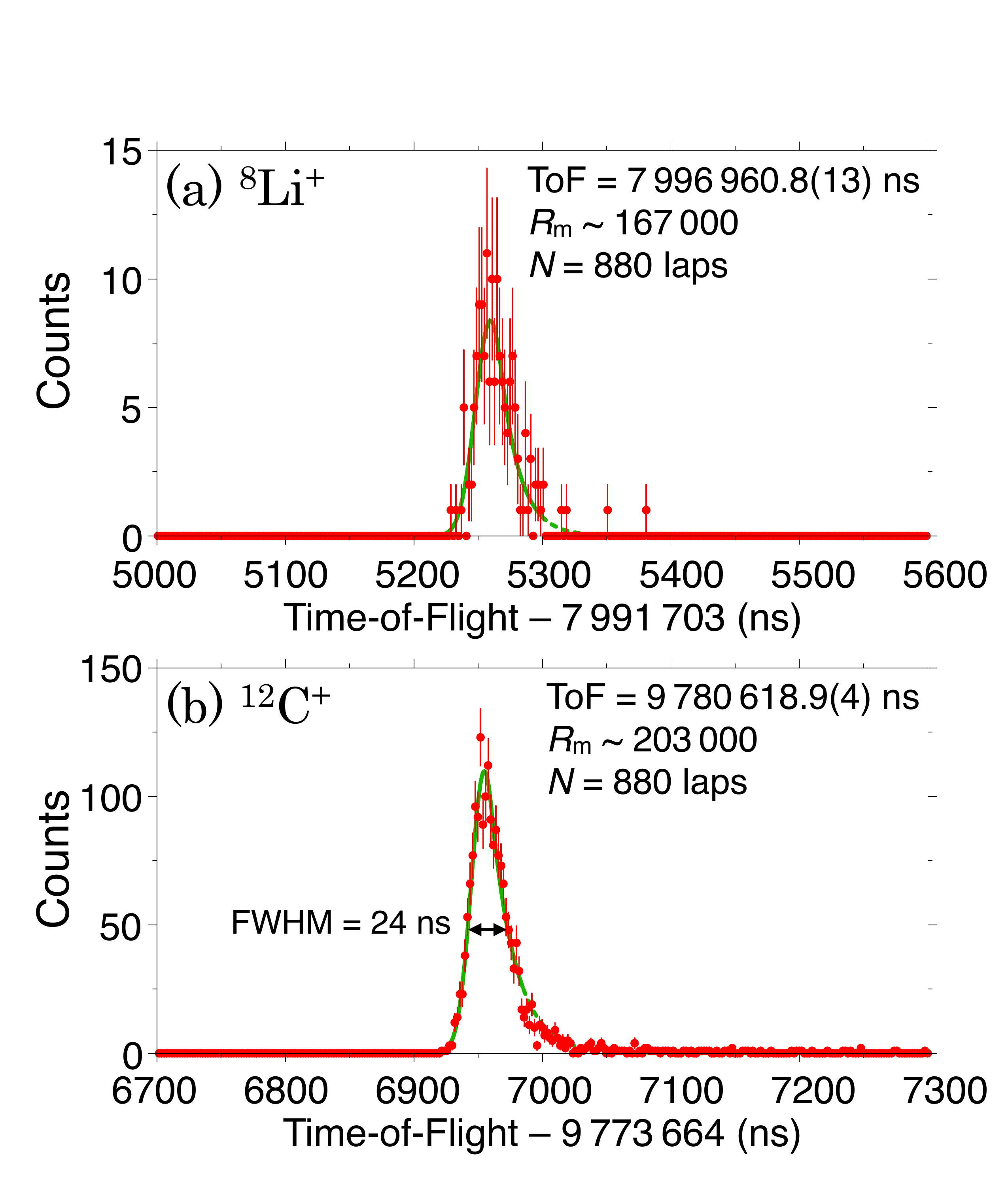}
  \caption{(Color online).
  Typical time-of-flight spectra for (a) $^{8}$Li$^{+}$ and (b) $^{12}$C$^{+}$ after 880 laps in the MRTOF.
  The slow tail is a result of higher-order ion optical aberrations in the mirrors and low-angle scattering from the residual gas in the reflection chamber.
  The fit shown is from Eq.~(\ref{eq:fit_func_001}), see text for details.
  }
  \label{fig:peak_fitting_both_001}
\end{figure}

 Typical time-of-flight spectra for $^{8}$Li$^{+}$ and $^{12}$C$^{+}$ are shown in Fig.~\ref{fig:peak_fitting_both_001} with times-of-flight of $t_8 \approx$ 8 ms and $t_{12} \approx$ 9.8 ms, respectively.
The spectra of  $^{8}$Li$^{+}$ were each accumulated for 600 s, while each spectrum of $^{12}$C$^{+}$ only required 50 s.
Mass resolving powers of $R_{m} \sim$~167\,000 for $^{8}$Li$^{+}$ and $R_{m} \sim$~203\,000 for $^{12}$C$^{+}$ were achieved.
The statistical uncertainty of ToF and thus of the mass of interest is given as
\begin{equation}
 \left( \delta m/m \right)^{\rm sta} = \alpha/\left( R_{m}\sqrt{N_{\rm ion}} \right),
 \label{eq:mass_precision_001}
\end{equation}
where $\alpha$ is close to unity for MRTOF measurements, $R_{m}$ is the mass resolving power and $N_{\rm ion}$ is the number of ions in the peak.
In principle, to perform the measurement the minimal ion number for the MRTOF is $N_{\rm ion} \ge 1$, as it is a true spectrograph, while generally PTMS requires $N_{\rm ion} \gtrsim 100$ to fit the resonance curve.
\par Due to higher-order ion optical aberrations in the mirrors and low-angle scattering from the residual gas during flight, the spectrum has a slow tail.
In order to properly take the tail into account, a Gaussian fitting function with an exponential-tail, as described by Eq.~(\ref{eq:fit_func_001}), was used \cite{Koskelo1981}.
 \begin{eqnarray}
 f(t) = 
 \left\{
 \begin{array}{l}
 A\ e^{-(t-t_{m})^{2}/2\sigma^{2}} \hspace{5mm} {\rm for}\ t\leq t_{m}+t_{c} \\
 \vspace{-2mm}\hspace{58mm} ,\\
 A\ e^{t_{c}(2t_{m}-2t+t_{c})/2\sigma^{2}} \hspace{5mm} {\rm for}\ t\geq t_{m}+t_{c}
 \end{array}
 \right.
 \label{eq:fit_func_001}
 \end{eqnarray}
where $A$ is the Gaussian peak height, $t_{m}$ is the Gaussian centroid, $\sigma$ is the standard deviation of the Gaussian, and $t_{c}$ is the distance from $t_{m}$ to the exponential tail switching point.
The shape parameters $t_{c}$ and $\sigma$ were determined from a high-statistics $^{12}$C$^{+}$ spectrum and fixed for all fittings.
\par To compensate for ToF drift caused by slight drifts of the MRTOF potentials over time, measurements of $^{8}$Li$^{+}$ interleaved those of $^{12}$C$^{+}$.
The effective ToF of the $^{12}$C$^{+}$ references were determined by linear interpolation of measurements before and after each $^{8}$Li$^{+}$ measurement.
\par In principle, the relation between the mass and the ToF is given by
\begin{equation}
 t = a \sqrt{m} + t_{0},
 \label{eq:tof2mass_001}
\end{equation}
where $a$ is a characteristic constant and $t_{0}$ is a constant time offset caused by the delay between the TDC start signal and the actual ejection of the ions from the ion trap.
Using picked up switching noise from trap ejection switch as a TDC stop signal, it was possible to measure the delay, which was found to be $t_{0} = 199$~ns.
However, to account for systematic uncertainty in the propagation path of the switching noise, a value of $\delta t_{0}^{\rm sys}~=~10$~ns was adopted.
Generally, to avoid systematic uncertainties from $t_{0}$, time-of-flight mass measurements interpolate between a pair of reference species.
In the case of MRTOF, due to the large ToF compared to $\delta t_{0}^{\rm sys}$, a single reference method can be used.
\par The mass of $^{8}$Li$^{+}$ is given by
\begin{equation}
  m_{8} = \left( \cfrac{t_{8}-t_{0}}{t_{\rm 12}-t_{0}} \right)^{2} m_{12} = \rho^{2} m_{12}.
 \label{eq:mass_calib_001}
\end{equation}
The statistical uncertainties $\delta m^{\rm sta}$ were determined from uncertainties derived from the ToF fittings, while $\delta t_{0}^{\rm sys}$ leads to a systematic uncertainty.
An expansion of Eq.~(\ref{eq:mass_calib_001}) up to the 1st order in ($t_{0}/t_{12}$) yields
\begin{equation}
m_{8} = m_{12}\left(\frac{t_8}{t_{12}}\right)^2 + 2m_{12}\frac{t_{8}(t_{8}-t_{12})}{t_{12}^3} t_{0}.
 \label{eq:mass_taylor_001}
\end{equation}
The effect of the uncertainty in $t_{0}^{\rm sys}$ is determined by the second term in Eq.~(\ref{eq:mass_taylor_001}).
In the case of $^{8}$Li$^{+}$, our adopted value of $\delta t_{0}^{\rm sys} = 10$~ns results in a systematic uncertainty of 3.4 keV.
\begin{table}[t]
 \centering
 \caption{
 $\rho^{2}$-values for $^{8}$Li$^{+}$, $^{7}$Li$^{+}$ and $^{9}$Be$^{+}$ along with the derived and literature mass excesses, $\Delta$.
 }
 \begin{tabular}{cccccll} \hline \hline
  Isotope & & $\rho^{2}$ & & $\Delta_{{\rm MRTOF}}$ (keV) & & $\Delta_{{\rm Lit}}$ (keV) \cite{Audi2012}  \\ \hline
  $^{8}$Li & & 0.668\,525\,53(14) & & 20\,947.6(15)(34) & & 20\,945.80(5) \\
  $^{7}$Li & & 0.584\,648\,35(77) & & 14\,911.4(9)(41) & & 14\,907.105(4) \\
  $^{9}$Be & & 0.751\,004\,25(22) & & 11\,352.6(25)(26) & & 11\,348.45(8) \\ \hline \hline
 \end{tabular}
 \label{tab:masses_001}
\end{table}
\par
\begin{figure}[t]
 \centering
  \includegraphics[bb=0 14 842 711, width=86mm]{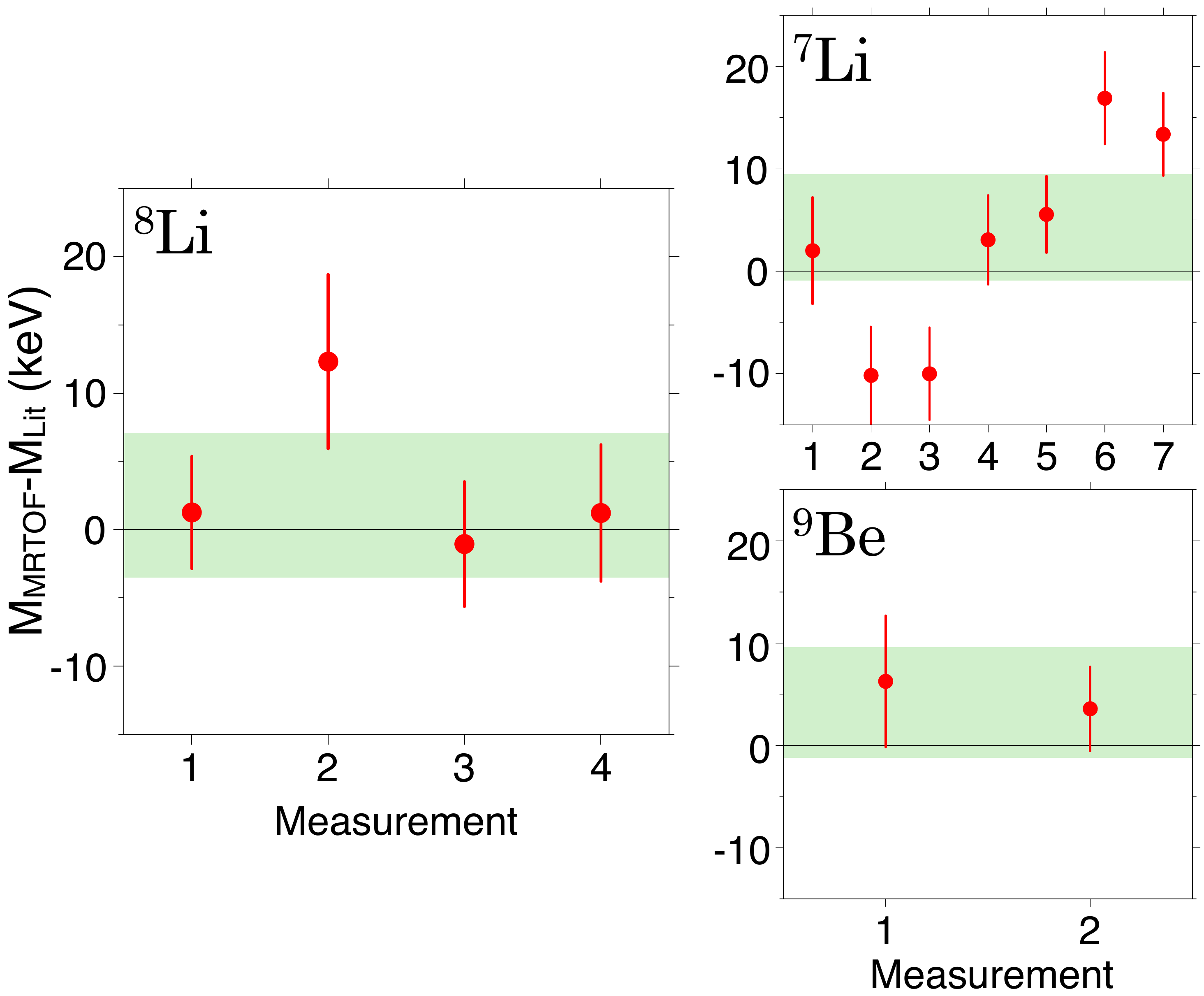}
  \caption{(Color online).
  Deviations of each measurements from AME2012 values \cite{Audi2012}.
  In each uncertainties the systematic uncertainties are included.
  The weighted averages with the systematic uncertainties are also shown by the green band.
  }
  \label{fig:8Li_mass_001}
\end{figure}

 To confirm this single reference method, the masses of $^{7}$Li$^{+}$ and $^{9}$Be$^{+}$ were similarly determined offline.
In all cases, the results were in agreement with the literature values.
The derived mass excesses are shown in Table~\ref{tab:masses_001} while Fig.~\ref{fig:8Li_mass_001} shows the individual deviations from the literature values.
The weighted average with a systematic uncertainty is represented by the green band.
The weighted average deviation of $^{8}$Li was found to be $\Delta m =$~1.8(15)(34)~keV, corresponding to a relative mass uncertainty of $\delta m/m = 6.6\times10^{-7}$.
The weighted average deviations of $^{7}$Li and $^{9}$Be were similarly evaluated and found to be $\Delta m =$~4.3(9)(41)~keV and $\Delta m =$~4.2(25)(26)~keV, respectively.
\par We have performed first online mass measurements with an MRTOF, using only $^{12}$C$^{+}$ references for the radioactive $^8$Li$^+$.  A mass resolving power of $R_{m} \sim$~167\,000 was achieved within 8 ms for $^8$Li$^+$, equivalent to a Penning trap with an 11 T magnetic field strength.  An accurate result was achieved with a relative mass uncertainty of $\delta m/m = 6.6\times10^{-7}$.
We verified the single reference method with $^{7}$Li$^{+}$ and $^{9}$Be$^{+}$.
\par This $^{8}$Li$^{+}$ measurement truly represents a worst-case scenario for the MRTOF.  The low mass-to-charge ratio minimizes the speed gain of the MRTOF over conventional PTMS.
In addition, the large fractional mass difference between $^{12}$C$^{+}$ and $^{8}$Li$^{+}$, which would not occur in any measurement of heavier nuclei, creates a maximally large systematic uncertainty from the ToF offset.
However, let's compare this to PTMS measurement of light-mass nucleus $^{11}$Li ($T_{1/2} = 8.75$ ms) \cite{Smith2008}.
The mass resolving power achieved in Ref.~\cite{Smith2008} was only $m/\Delta m~=~86\,000$, with an excitation time of 18 ms, however the total number of detected $^{11}$Li approached $N~=~10\,000$.
Were our MRTOF utilized under similar conditions, a resolving power of $m/\Delta m~\approx~170\,000$ could be achieved in 9.4 ms, resulting in less decay loss and faster accumulation of statistics.
With $\delta t_{0}^{\rm sys}$~=~10~ns, the precision limit using $^{12}$C as a reference would be $(\delta m/m)^{\rm sys}~=~9.5\times10^{-8}$, however $^{11}$B$^{+}$ from laser ablation as a reference would yield $(\delta m/m)^{\rm sys}~=~3.3\times10^{-9}$.
Based on Eq.~(\ref{eq:mass_precision_001}), the statistical uncertainty limit would be $5.9\times10^{-8}$ with a similar $N_{\rm ion}$ which is competitive with the relative mass uncertainty of $\delta m/m~=~6.2\times10^{-8}$ achieved in Ref.~\cite{Smith2008}.
Considering decay losses, the MRTOF could actually achieve better relative uncertainty than PTMS.
\par With the very light $^{8}$Li$^{+}$, we achieve mass resolving powers competitive with conventional PTMS of short-lived nuclei by using shorter observation times.
We have verified the speed, precision and accuracy of the technique online.
For short-lived, heavy nuclei such as trans-uranium nuclei and nuclei important to r-process nucleosynthesis we believe this new method will truly be a boon.
We plan to begin measurements of trans-uranium elements and of r-process nuclei in FY2013.


\begin{acknowledgments}
This experiment was carried out under Program No.\ NP0702-RRC01-25 at the RIBF operated by RIKEN Nishina
Center.
This work was supported by the Japan Society for the Promotion of Science KAKENHI (Grant Nos.\ 2200823, 24224008 and 24740142).
We wish to gratefully acknowledge the RIBF accelerator crew and the RIKEN Accelerator Research Facility for their support.
\end{acknowledgments}

\bibliography{ito_8Li_paper_proof_v1}


\end{document}